\documentclass[
reprint,
aps,
twocolumn,
superscriptaddress,
longbibliography]{revtex4-1}

\usepackage{bm}
\usepackage{amssymb}
\usepackage{amsfonts}
\usepackage{amsmath}
\usepackage{graphicx}
\usepackage{theorem}
\usepackage{color}
\usepackage{soul}
\usepackage{todonotes}
\usepackage[normalem]{ulem}

\usepackage{hyperref}
\usepackage{xcolor}
\hypersetup{
	colorlinks,
	linkcolor={blue!50!black},
	citecolor={blue!50!black},
	urlcolor={blue!80!black}
}

\newcommand{\brk}[1]{\left(#1\right)}

\newcommand{\matrixII}[4]{\left(\begin{array}{cc}#1&#2\\#3&#4\end{array}\right)}

\newcommand{\figref}[1]{Fig.~\ref{#1}}

\newcommand{\Eqref}[1]{Eq.~(\ref{#1})}

\newcommand{\bbar}{\bar{\mathfrak{b}}}

\renewcommand{\a}{\mathfrak{a}}
\newcommand{\abar}{\bar{\mathfrak{a}}}

\renewcommand{\b}{\mathfrak{b}}

\renewcommand{\d}{\mathbf{d}}

\newcommand{\nablaa}{{\nabla}^\a}
\newcommand{\nablabar}{{\nabla}^{\abar}}

\begin{document}

\title{Geometric approach to mechanical design principles in continuous elastic	 sheets}

\author{Michal Arieli} 
\author{Eran Sharon}
\email{erans@mail.huji.ac.il}
\author{Michael Moshe}
\email{michael.moshe@mail.huji.ac.il}
\affiliation{Racah Institute of Physics, The Hebrew University of Jerusalem, Jerusalem, Israel 91904}

\begin{abstract}
Using a geometric formalism of elasticity theory we develop a systematic theoretical method for controlling and manipulating the mechanical response of slender solids to external loads. 
We formally express global mechanical properties associated with non-euclidean thin sheets, and interpret the expressions as inverse problem for designing desired mechanical properties. 
We show that by wisely designing geometric frustration, extreme mechanical properties can be encoded into a material using accessible experimental techniques. 
To test the methodology we 
derive a
family of geometries that 
result with
anomalous mechanical behavior such as tunable, an-harmonic, and even vanishing rigidities. The presented formalism can be discretized and thus opens a new pathway for the design of both continuum and discrete solids and structures.
\end{abstract}
\maketitle

\textit{Introduction} The study of non conventional mechanics of solid structures has attracted a lot of attention in recent years. Negative elastic moduli (e.g. Poisson's ratio) \cite{Bertoldi2010Negative}, controllable instabilities \cite{Florijn2014Programmable}, and topologically protected directed edge states \cite{Kane2014topological, Paulose2015topological} are 
representative
unusual mechanical properties of discrete structures composed of rigid or deformable elements \cite{Bertoldi2017NatureReview},
realized for example in Origami, Kirigami, and 2D Maxwell lattices\cite{Mao2018maxwell}.

The typical controllable degrees of freedom in discrete structures are the building blocks' defining parameters and their spatial arrangement.
For example, effective elastic moduli and instabilities can be controlled by twisting triangles in the Kagome lattice \cite{sun2012surface, fruchart2020dualities}, by patterning holes sizes and structure in deformable metamaterials \cite{shim2015harnessing}, by pruning bonds in disordered solid models \cite{goodrich2015principle}, or by introducing topological defects into complex rigid mechanical metamaterials \cite{meeussen2020topological}.
An immediate question then is whether analogous design principles exists for continuous matter, and if so, what  the natural controllable degrees of freedom are? 
One possible answer comes from the study of thin sheets that undergo controllable non-uniform inelastic deformations
such as actuatable responsive gels \cite{Klein2007,Kim2012}, sheets made of nematic elastomers  \cite{ware2015voxelated} and electro-active polymers \cite{hajiesmaili2019reconfigurable}.
In these systems non-uniform 'growth' fields can be induced within the sheets, determining their intrinsic geometry in a universal way.

The growth profile averaged across the thickness sets the sheets' lateral reference lengths, and its variation across the thickness determines its reference curvature, which forms an independent controllable geometric degree of freedom. The preferred lengths and curvatures are quantified in terms of a metric tensor $\abar$ and extrinsic-curvature tensor $\bbar$, correspondingly. 
Upon inscribing nonuniform reference fields on a thin elastic sheet  it typically deforms into a 3d shape, selected by minimizing an elastic energy, which form an additive measure for local deviations of actual lengths and curvatures from their reference values \cite{Efrati2009,Koiter1966}. Therefore, $\abar$ and $\bbar$ are, in fact, controllable degrees of freedom of continuum thin sheets, that can affect their geometry and, possibly, their mechanics.   
Due to its fundamental importance, a plethora of techniques have been developed  
for determining $\abar$ and $\bbar$, including controlled injection \cite{Klein2007}, surface patterning \cite{ware2015voxelated}, channels patterning in inflatable sheets \cite{siefert2019bio}, selective cross-linking  \cite{wu2013three}, 3D printing  \cite{gladman2016biomimetic} and sewing \cite{henderson2001crocheting}.
Extensive experimental and theoretical studies were dedicated to incompatible sheets, in which the inscribed reference fields are controlled independently,
and cannot be realized simultaneously by any admissible surface in 3d Euclidean space. Consequently, the selected configuration is geometrically frustrated, and it minimizes an elastic energy that compromises between metric and curvature discrepancies. 
{Manipulation of $\abar$ and $\bbar$ was successfully applied in order to achieve non-trivial 3-d configurations, i.e., non-trivial ground state \cite{aharoni2018universal}. However, there is no systematic study of how to manipulate these fields in order to achieve non-trivial, desired mechanical properties, i.e., shaping the energy landscape around the ground state.}

In this work we develop a theoretical framework in which the reference metric and curvature fields $\abar$ and $\bbar$ are unknowns that are found by constraining the elastic energy to match desired compliance against external loads. Upon deriving a formal equation that relates the reference fields with the emerging elastic properties,
we solve a prototypical inverse-problem and discover a rich family of frustrated non-euclidean ribbons whose  energy is either weakly or strongly degenerate under deformations, and thus forming ``anomalously'' soft springs.

\textit{Theoretical Framework} Within a geometrical formalism of elasticity, the state of an elastic sheet is characterized in terms of  metric tensor $\a$ and curvature tensor $\b$, measuring lengths and curvatures in the actual configuration.
A given configuration with $\a$ and $\b$ may differ from the reference fields $\abar$ and $\bbar$. The elastic energy density then quantifies local deviations from these reference fields and consists of a stretching term $\mathcal{W}_\text{st}(\a;\abar)$ and a bending term $\mathcal{W}_\text{bn}(\b;\bbar)$ that measure metric and curvature discrepancies, with the total energy of the form 
\begin{align}
E(\a,\b;\abar,\bbar) = \int_{\Omega} \d S_{\abar} \brk{ \mathcal{W}_\text{st}(\a;\abar) +  \mathcal{W}_\text{bn}(\b;\bbar)}.
\label{eq:Energy}
\end{align}
A key point in the theory of non-euclidean sheets that lies at the heart of our work is that unlike the actual fields $\a$ and $\b$, the reference fields $\abar$ and $\bbar$ are in principle independent and can lead to an elastic sheet that cannot simultaneously satisfy both reference fields \cite{Armon2011,meiri2021cumulative}. 

{Since a 3d configuration has three degrees of freedom at every point, }
the actual fields describing an admissible surface, which consists of six degrees of freedom, cannot be independent but must satisfy a set of compatibility conditions aka Gauss-Peterson-Mainardi-Codazzi equations (GPMC)
\begin{align}
K(\a) = \det\brk{\a^{-1}\b},\quad 
\nablaa_\alpha \b_{\beta\gamma} =  \nablaa_\beta \b_{\alpha\gamma}.
\label{eq:GPMC}
\end{align}
Here $K(\a)$ and $\nablaa$ are the Gaussian curvature and covariant derivative associated with the actual metric $\a$. 	
The equilibrium configuration is obtained by minimizing the elastic energy in \Eqref{eq:Energy} with respect to $\a,\b$, subjected to \Eqref{eq:GPMC}.
Geometric incompatibility (and consequently frustration) occurs when either of the GPMC equations are violated by $\abar$ and $\bbar$ \cite{siefert2021euclidean}, in which case no admissible configuration can satisfy $\abar$ and $\bbar$ simultaneously, and the elastic energy in \Eqref{eq:Energy} at the ground-state is finite.


The task of analytically solving for the energy minimizing configuration, either in the presence or absence  of external constraints, is in general quite cumbersome. However, the theory significantly simplifies for the important class of very thin elastic sheets
, dominated by stretching energy and therefore $\a \to \abar$ with $\mathcal{W}_\text{st}\left(\a,\abar\right) \to 0$. 
In this work we consider this asymptotic limit and thus assume $\a = \abar$. In this limit the remaining unknown is $\b$ with respect to which  bending energy should be minimized subjected to 
\begin{align}
K(\abar) = \det\brk{\abar^{-1}\b},\quad 
\nablabar_\alpha \b_{\beta\gamma} =  \nablabar_\beta \b_{\alpha\gamma}.
\label{eq:beq}
\end{align}

Note that even in the very thin limit, if external constraints are incompatible with $\abar$, e.g.  very large tensile forces acting on the boundary, the actual metric $\a$ may deviate from its reference value. This type of deformations, however, are not generic, and the common response of a very thin sheet to external loads are isometric $\a=\abar$, and thus avoids stretching energy. Consequently, the mechanical properties of a thin sheet to external loads reflects the energetic cost of bending.

Isometric deformations are described by the solution to \Eqref{eq:beq}: upon solving this set of differential-algebraic equations, there are free integration variables formally denoted by $\phi$, that parametrize the space of admissible configurations satisfying $\a = \abar$, which can be of finite or infinite dimension. Global deformations are naturally expressed in terms of the configurational variables $\phi$, and different flows in $\phi$-space represent different isometric deformations. After substituting  the solution of \Eqref{eq:beq} for $\b(\phi)$ in the energy functional it takes the form $E_{\abar,\bbar} (\phi)$, with the ground state $\phi_\text{gs}$ minimizing this energy.


An important outcome of the framework described here is that the resistance of a thin sheet to an isometric deformation described by $\phi(s)$ is
\begin{align}
\kappa_\text{eff} = \frac{1}{2}\frac{\partial^2 E_{\abar,\bbar}(\phi(s))}{\partial s^2}|_{\phi_\text{gs}}.
\label{eq:rigidity}
\end{align}

The main message of this paper is that the energy landscape $E_{\abar,\bbar}(\phi(s))$ can be manipulated via control of the reference fields $\abar$ and $\bbar$. While it is not yet clear to what extent the energy landscape can be designed across all configurations space, in this work we focus on designing the energy landscape at the vicinity of the ground-state. In particular, \Eqref{eq:rigidity} is interpreted as an equation in which $\kappa_\text{eff}$ is a given desired mechanical property with the reference fields $\abar$ and $\bbar$ taken as unknowns.  This define the inverse problem that is the main objective of our work: 
\begin{quote}
	Given a desired rigidity for ribbon deformation  $\kappa_\text{eff}$, find $\abar$ and $\bbar$ that satisfy \Eqref{eq:rigidity} 
\end{quote}
%
 Since  $\abar$ and $\bbar$ consist of six independent degrees of freedom \Eqref{eq:rigidity} is redundant, which implies that with wisely selected $\abar$ and $\bbar$ extreme mechanical properties can be encoded into a thin sheet.  

Before solving an inverse problem we start with a forward one that demonstrates the derivation of \Eqref{eq:rigidity} for an incompatible ribbon - a system that has been previously studied in various natural and synthetic contexts, and we show that $\kappa_\text{eff}$ can be manipulated by tuning $\bbar$. 

\textit{Geometrically Incompatible Ribbon}
One particularly simple technique for inscribing a reference curvature field $\bbar$  is by gluing two stretched thin sheets to form a bi-layer \cite{Armon2011}. In an appropriate coordinate system, the reference fields take the form 
\begin{equation}
\abar = \matrixII{1}{0}{0}{1} \quad \bbar = \matrixII{k_1}{0}{0}{k_2},
\label{eq:refGuest}
\end{equation}
with the principle preferred curvatures $k_{1,2} $ encoding the strains imposed on the two layers (see	\cite{Armon2011} for details).  

\begin{figure}
	\centering
	\includegraphics[width=\linewidth]{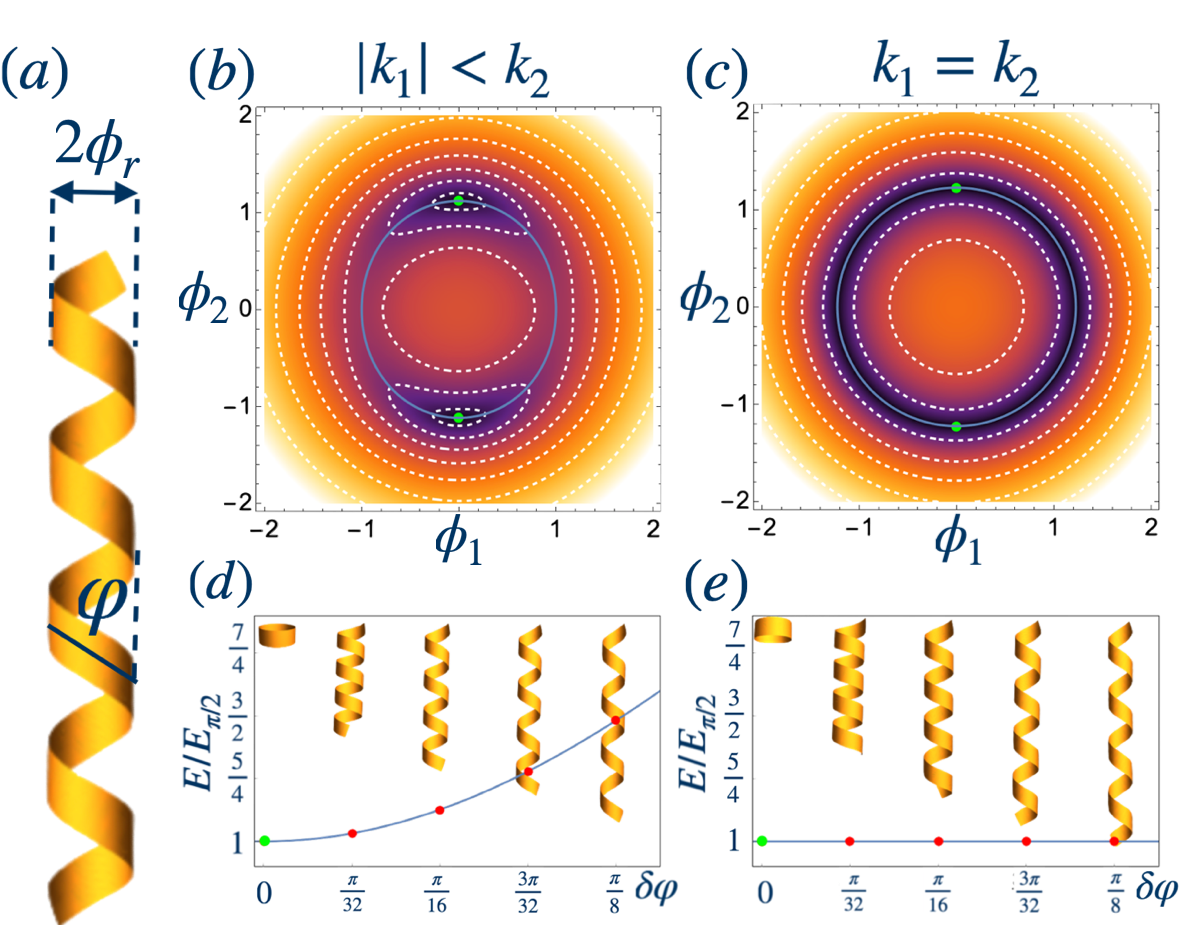}
	\caption{Energy landscapes and isometric configurations in very thin non-euclidean ribbons with the reference fields in  \Eqref{eq:refGuest}. (a) Isometric configuration quantified by $(\phi_r,\varphi)$ as in \Eqref{eq:PolarGuestCurvature}. (b-c) Energy landscape in configuration space for incompatible ribbon with a single (b) and degenerated (c) ground-states marked by a green dot. Pitch deformations are imposed by displacing  $\varphi =\pi/2 + \delta \varphi$ while leaving $\phi_r$ free, depicted by blue curves.  (d-e): Representative configurations along the deformation curve with their corresponding energy values normalized by the ground-state energy.}
	\label{fig:figure1}
\end{figure}

The Gaussian curvature associated with $\abar$ and $\bbar$ are $K(\abar) = 0$ and $K(\bbar) = k_1 k_2$, hence the first equation in  \Eqref{eq:beq} is violated and the elastic sheet is geometrically frustrated. 
In the limit of a thin ribbon $t \ll w \ll  \ell$, with $t$ the thickness, $w$ the width, and $\ell$ the length, \Eqref{eq:beq} constrains the curvature to satisfy $\det\brk{\b} = 0$.
Furthermore, since the ribbon is uniform along its long direction, we assume the actual fields can vary at most across the width, in which case the solution of \Eqref{eq:beq} for $\b$ reads 
\begin{equation}
\b = \matrixII{\phi_1^2}{-\phi_1\, \phi_2}{-\phi_1\, \phi_2}{\phi_2^2}.
\label{eq:GuestCurvature}
\end{equation}
Here $\phi = (\phi_1,\phi_2)$ is spatially uniform and configuration space is two-dimensional. 
{To proceed and find the energy ground-state we have to specify the functional form of $\mathcal{W}_\text{st}$ and $\mathcal{W}_\text{bn}$ in \Eqref{eq:Energy}. While the results in this work are applicable to any positive definite energy functional, here we 
follow the simplest model of a quadratic (i.e. Hookean) energy density  \cite{Efrati2009}
\begin{equation}
\begin{split}
	\mathcal{W}_\text{st} &= \frac{t}{8} \mathcal{A}^{\alpha\beta\gamma\delta} (\a - \abar)_{\alpha\beta}(\a - \abar)_{\gamma\delta} \\
	\mathcal{W}_\text{bn} &= \frac{t^3}{24} \mathcal{A}^{\alpha\beta\gamma\delta} (\b - \bbar)_{\alpha\beta}(\b - \bbar)_{\gamma\delta} ,
\end{split}
\end{equation}
with  $\mathcal{A}^{\alpha\beta\gamma\delta}  = \tfrac{Y}{1-\nu^2} \left(\nu\, \abar^{\alpha\beta}  \abar^{\gamma\delta}  +(1-\nu)\abar^{\alpha\gamma}  \abar^{\beta\delta}\right)$ the elastic tensor with Poisson's ration $\nu$ and Young's modulus $Y$  \cite{Efrati2009}.
The energy minimizing configuration is}
\begin{eqnarray}
\phi_{gs} =  \left\{
\begin{array}{ll}
(0,\sqrt{\nu k_1 + k_2}) & |k_1| < k_2 \\
(\sqrt{k_1 + \nu k_2},0) & |k_2| < k_1
\end{array} 
\right.
\end{eqnarray}
The physical interpretation of $\phi = (\phi_1,\phi_2)$ is easily recognized in ``polar'' parametrization: $\phi = \phi_r(\cos(\varphi),\sin(\varphi))$, where \eqref{eq:GuestCurvature} takes the form
\begin{equation}
\b =  R(\varphi)^T \matrixII{0}{0}{0}{\phi_r^2} R(\varphi)
\label{eq:PolarGuestCurvature}
\end{equation}
and $R(\varphi)$ is the rotation operator.
The radial component $\phi_r$ measures the curvature magnitude, and $\varphi$ is the angle between the curvature principle direction and the ribbons orientation, as illustrated in \figref{fig:figure1}(a).
In this parametrization variations in $\varphi$ correspond to pitch deformations whereas variations in $\phi_r$ corresponds to twist deformations similar to torsional-springs.
{Without loss of generality we assume in \Eqref{eq:refGuest} that the principle curvatures satisfy $|k_1| \leq k_2$, in which case the energy has a minimum at $\phi_r = \sqrt{k_2+\nu k_1}$ and $\varphi=\pi/2$, that is the ribbon curls along the direction of the larger curvature to minimize energy. The ground-state and energy landscapes in configurations space are shown in \figref{fig:figure1}(b),(d) for $\nu=0.5$.  
If the ribbon length coincides with the dominant curvature orientation $k_2$, at the ground-state the ribbon rolls onto itself to form a ring shape, as shown in inset of \figref{fig:figure1}(d). A pitch deformation, which corresponds to pulling on the ribbons ends, is imposed by fixing $\varphi$ while minimizing energy with respect to $\phi_r$. The deformation trajectory in configuration space and the energy along it with representative configurations are shown in \figref{fig:figure1}(b),(d).}

{Since the actual curvature in \Eqref{eq:PolarGuestCurvature} is uniform the integration in \Eqref{eq:Energy} is trivial and we can calculate analytically any property of the system. Denoting the ribbon's area by $S_{\Omega}$} we find the system resistance against the two deformations modes  
\begin{equation}
\begin{split}
	\kappa_\text{eff}^\varphi &= \frac{Y S_\Omega}{3(1+\nu)} |k_2 - k_1| (\nu\, k_1+k_2) \\
	\kappa_\text{eff}^r &= \frac{2 Y S_\Omega}{3(1-\nu^2)}(\nu\, k_1+k_2) 
\end{split}
	\label{eq:GuestModuli}
\end{equation}
\Eqref{eq:GuestModuli} is an  explicit forms of \Eqref{eq:rigidity} for the case of geometrically frustrated ribbon. As we suggested in the context of \Eqref{eq:rigidity}, $k_1$ and $k_2$ (the principle reference curvatures) can be treated as unknowns while $\kappa_\text{eff} = (\kappa_\text{eff}^r,\kappa_\text{eff}^\varphi)$  are known desired elastic moduli. {It is clearly seen that the two rigidities can be tuned by a proper selection of $k_1$ and $k_2$}

\textit{Anomalous Ribbon}
\Eqref{eq:GuestModuli} shows that extreme mechanical properties can be designed. One extreme case corresponds to $k_2= - \nu k_1$, which leads to $\kappa_\text{eff}^r=0$ no resistance against torsional deformation. Since the analysis is valid only for $|k_1|\leq k_2$, and since $-0.5<\nu\leq 1$, this solution is valid only for $\nu=1$ and reduces to $k_2=-k_1$. This anomalous mechanical property is not generic as it requires specific material properties.

Another extreme case is a vanishing resistance to the pitch-like deformation, which loses its resistance if $k_1=k_2$. The degeneracy in this case, which was discovered in \cite{schenk2014zero}, has a simple root: The reference metric is euclidean which implies the actual Gaussian curvature in the thin limit must vanish. The extrinsic reference curvature is isotropic, hence, to preserve zero Gaussian curvature, the ribbon curves uni-axially along a spontaneously selected orientation. The pitch-like deformation does not modify the curvature magnitude, but only its orientation, and therefore costs no energy.
This is summarized in  \figref{fig:figure1} (c),(e) where the energy landscape in configurations space and pitch deformation trajectory is shown together with representative degenerated configurations.

\begin{figure*}
	\centering
	\includegraphics[width=1.\linewidth]{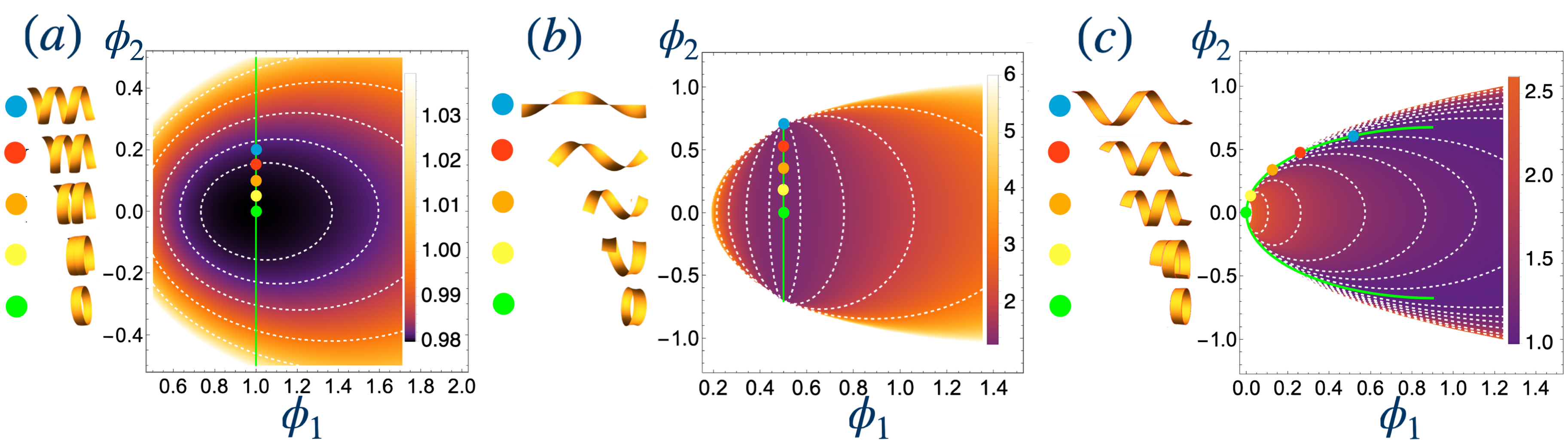}
	\caption{Energy landscape and representative configurations for incompatible ribbons with isotropic reference fields as in \Eqref{eq:refConf}, with the green dot representing the ground-state. (a) A ribbon with  $k_1(v) = k_2(v)$ presenting weak anomalous response corresponding to energy quartic in the deformation variable. (b) A ribbon with zero reference curvature $k_1(v) = k_2(v)=0$ presenting strong degeneracy reflecting the geometric property of Bonnet transformations. (c) A ribbon with uniform reference fields presenting strong degeneracy equivalent to the one presented in \figref{fig:figure1} (c)}
	\label{fig:Main-Figure}
\end{figure*}

\textit{Inverse problem}
A natural question that the {previous example} raises is whether 
other reference fields that
lead to anomalous mechanical response exist?
To address this question we define the notions of weak and strong energetic degeneracy:
Let $\phi^*$ be the energy ground-state and $\phi(s)$ a trajectory in configurations space with $\phi(0) = \phi^*$. A system is weakly degenerated if its linear rigidity vanishes $\partial_s^2 E(\phi(s))|_{s=0} = 0$. A system is strongly degenerated along $\phi(s)$ if the energy is constant along it $E(\phi(s)) = E_0$. 
The key step towards finding reference fields that lead to either type of degeneracy, or to other mechanical properties, is solving \Eqref{eq:beq} and substituting in the energy to express it in terms of $\abar,\bbar$.  Then, given a desired mechanical property, e.g. anomalous rigidity, we derive it from the energy and consider it as a constraint on the reference fields. 
In what follows we demonstrate the method for both weak and strong degeneracy.


Our strategy is to study the mechanical response for a wide yet limited family of reference fields. We note that in 2d geometry all metric tensors are conformally equivalent, that is, there always exists a coordinate system in which the reference metric takes the form $\abar_{\mu\nu} = e^{2\lambda} \, \delta_{\mu\nu}$. Since $\abar$ is isotropic the reference curvature $\bbar$ can always be diagonalized. 

{We, thus, study ribbons with conformal reference metric and a reference curvature in a Cartesian-like coordinate system $(u,v)$. 
We focus on reference fields that depend only on one coordinate, $v$, along the ribbon.
}
\begin{equation}
\abar = \lambda(v)^2 \matrixII{1}{0}{0}{1} \quad \bbar = \matrixII{k_1(v)}{0}{0}{k_2(v)},
\label{eq:refConf}
\end{equation}
The solution of \Eqref{eq:beq} that is invariant along the ribbon
\begin{equation}
\b = 
\matrixII{h}{\phi_2}{\phi_2}{\tfrac{\phi_2^2 + \lambda'(v)^2 - \lambda(v) \lambda''(v)}{h}	}.
\label{eq:ConformalCurvature}
\end{equation}
Here $h = \sqrt{ \phi_1 \lambda(v)^2  -  \phi_2^2 - \lambda'(v)^2}$ and configurations space is two dimensional  and spanned by $\phi = (\phi_1,\phi_2)$. 

Next step is to find an energy minimizer $\phi^*$ and study the mechanical response to configurational perturbations. 
While in principle this task is possible, analytic progress is daunting due to the generality of the reference fields $\abar$ and $\bbar$.  An important scenario in which the analysis is simplified is where anomalous rigidity holds not only globally at the level of total energy, but also locally for each material element at the level of energy density. Upon requiring this property we find that energy-density gradient identically vanishes when $\abar \propto \bbar$, i.e. $\abar$ and $\bbar$ are isotropic.
In this scenario there are three possible cases:

Case I - {The general form: } 
In this case $k_1(v) = k_2(v) \equiv k(v)$ in \Eqref{eq:refConf} and for $k'(v)\neq 0$ the energy gradient vanishes for $\phi_1^* = \frac{2 \lambda'(v)^2 - \lambda(v) \lambda''(v)}{\lambda(v)^2}$ and $\phi_2^* = 0$. Since $\phi_1^*$ is constant, the function $\lambda(v)$ is constrained to satisfy a differential equation whose solution is $\lambda(v) = \lambda_0 \, \text{sech}\brk{c (v - v0)}$
where $\lambda_0$ and $v_0$ are constants of integration, and $c$ is an arbitrary constant that sets the energy minimizing configuration $(\phi_1,\phi_2) = (c^2,0)$. The minimum energy is obtained for a point in configurations space and strong degeneracy is inaccessible in general. We calculate the rigidity at the ground state from a second order expansion of the energy and we find that resistance to configurational deformations is identical in all direction
\begin{equation}
\kappa_\text{eff} \propto 2 \lambda(v)^2 c  - (1+\nu) \lambda_0 k(v).
\label{eq:weakdeg}
\end{equation}
To achieve anomalously soft response we take
\begin{equation}
k(v) = \tfrac{2 \lambda(v)^2 c }{(1+\nu) \lambda_0}.
\end{equation}
In this case {resistance is equally soft in all directions in configurations space, yet the degeneracy is weak:} the lowest order term in the elastic energy is quartic, leading to a weak anomalous response in a an-harmonic spring.
The energy landscape in configurations space and the configurations along the  $\phi_2$ direction are presented in \figref{fig:Main-Figure}(a). 

Case II - {Vanishing $\bbar$.} {In this case $k_1 = k_2 = 0$ in \Eqref{eq:refConf} and the energy gradient is 
\begin{equation}
\begin{split}
\nabla_\phi E = \int_{\Omega} \d S_{\abar} \brk{\lambda''(v) - \phi_1 \lambda(v)} (F_1,F_2),
\end{split}
\end{equation}
with $\nabla_\phi E \equiv (\partial_{\phi_1}E,\partial_{\phi_2} E)$ the gradient in configurations space and $F_1, F_2$ functions of $\phi_1, \phi_2$ and $\lambda(v)$ and its derivatives.
The requirement for energy extremum $\nabla_\phi E = 0$ results, in principle, with two equations for $(\phi_1,\phi_2)$. Since we enforce the gradient to vanish locally we find $\phi_1^* =  \lambda''(v)/ \lambda(v)$.
Since $\phi_1^*$ is a constant, our requirement is admittable  only if  $\lambda''(v)/ \lambda(v) = c^2$, which selects $\abar$ and is equivalent to}
\begin{equation}
\lambda(v) = \lambda_0 \cosh(c (v- v0)).
\label{eq:Ido}
\end{equation}
At the minimum a ribbon of width $w$ has an energy $E = \frac{2 c}{1+\nu} \frac{\tanh \brk{c\, w}}{w} S_{\Omega}$ independent of $\phi_2$, and thus presents a strong degeneracy to deformations along $\phi_2$. The energy landscape in $\phi$ space and the configurations along the degenerated curve $\phi^*_1 = c^2$ is presented in \figref{fig:Main-Figure}(b).
This anomalously soft ribbon is different from the one observed in right panel of \figref{fig:figure1}: It reflects a geometric property of the degenerated ground-states forming a family of isometric minimal surfaces related to each other by what is known as Bonnet transformation \cite{do2016differential}, and was studied experimentally in \cite{levin2016anomalously}.

Case III - {Uniform $\bbar$.} {In this case $k_1 = k_2 \equiv k $ and $\lambda$ are constants. The energy gradient in $\phi$ space is then
\begin{equation}
\begin{split}
\nabla_\phi E = \int_{\Omega} \d S_{\abar} \brk{\lambda^2  \phi_2 - k \sqrt{\lambda^2  \phi_2 - \phi_1^2} (1+\nu)} (G_1,G_2),
\end{split}
\label{eq:}
\end{equation}
where $G_1$ and $G_2$ are constants that depend on $\phi_1,\phi_2$ and $\lambda$. Similar to case II, 
since we require local anomalous rigidity we enforce $\lambda^2  \phi_2 - k \sqrt{\lambda^2  \phi_2 - \phi_1^2}(1+\nu) = 0$. Since this condition is under-determined it describes a continuous set of points at which the energy is in extremum. A direct analysis confirms that this curve describes a degenerated set of energy minimizers.}
	
The energy landscape in $\phi$ space and the configurations along the degenerated curve are presented in \figref{fig:Main-Figure}(c). 
We note that the reference fields obtained in this case are related with those of the anomalous ribbon in right panel of \figref{fig:figure1} via isometric parametrization of the $\phi$ parameters, hence the two systems are equivalent.

\textit{ Global vs. local mechanical degeneracy\\}
So far we required a point-wise degeneracy, i.e., the anomalous response holds for each material element regardless of the ribbon's  dimensions. However, anomalous response would also hold if the total energy is preserved under deformation, that is a global condition.

To demonstrate the effect of requiring global mechanical degeneracy we refine case III of isotropic uniform reference fields to account for non-uniform reference curvature. In this case the reference curvature $\bbar$ and the solution of \Eqref{eq:beq} for the actual curvature $\b$  are
\begin{equation}
\bbar = 
\matrixII{k(v)}{0}{0}{k(v)} ,\quad  \b = \matrixII{\phi_1}{\sqrt{\phi_1 \phi_2}}{\sqrt{\phi_1 \phi_2}}{\phi_2} .
\label{eq:bGlobal}
\end{equation}
Upon calculating the gradient of bending energy (per unit length) in $\phi$ space we find the condition for energy minimization in the form 
\begin{equation}
\int \d v \brk{\phi_1 + \phi_2 - (1+\nu) k(v)} = 0
\label{eq:GradGlobal}
\end{equation}
Global degeneracy occurs along the curve in configurations space defined by $\phi_1 + \phi_2 = (1+\nu)\left<k(v)\right>$ where $\left<\cdot\right>$ stands for integrating along the $v$ coordinate. In this case the energy is strongly degenerated as can be seen from its explicit form when evaluated along the ground-state curve:
\begin{equation}
E = \tfrac{\kappa_B}{1-\nu} \int \d v \brk{k(v)^2 + \brk{k(v) - (1+\nu) \left<k(v)\right>}^2}
\end{equation}
with $\kappa_B$ the bending rigidity. We conclude that there is an infinite set of independent reference curvatures that vary in space along the $v$ direction and lead to strong mechanical degeneracy. Similar refinement for global degeneracy holds as well for case II. 

\textit{Summary and discussion}
In this work we showed that the reference geometric fields $\abar$ and $\bbar$ that quantify reference distances and curvatures between adjacent material elements in an elastic sheet form a basic controllable degrees of freedom for designing the elastic energy landscape. Using this approach we showed that the geometry of the energy landscape at the vicinity of the ground-state can be controlled and extreme mechanical response can be programmed into the material. Specifically, we showed that upon controlling experimentally accessible reference fields $\abar$ and $\bbar$, strong and weak energy degeneracy can be programmed into an elastic sheet, either locally or globally.

The results of our work complement previous approaches and methods for controlling mechanical properties of 2d discrete structures of masses and springs which form a common model of matter. In that case the controllable degrees of freedom are the network connectivity and the reference values of spring lengths. Previous works studied network's connectivity as the controllable property to program mechanical response. A discrete realization of our work allows to program a discrete  structure of masses and springs via control of reference values of spring lengths and curvature.

Our work raises an interesting theoretical question: Assume that we can impart any desirable reference field on an elastic sheet, that is, we can control six independent degrees of freedom at each point. Configuration space is three dimensional at each point, i.e. of lower dimensionality. What are the limitations on sculpting energy landscape via manipulation of the reference fields? The answer to this question will necessarily depend on the precise form of the selected energy functional.

To enable analytical progress, in the current work we focused on highly symmetric systems. This allowed a direct calculation of the desired reference fields. A future promising direction is solving such inverse problems using numerical methods, which will allow studying systems of lower symmetry.



\begin{acknowledgments}

	MM acknowledges support from the Israel Science Foundation (grant No. 1441/19).
\end{acknowledgments}

%
%
%
%
%

%


\bibliography{references}  

\end{document}